\documentclass{spie}
\usepackage{graphicx} 
\usepackage{authblk}
\usepackage[export]{adjustbox}
\usepackage[verbose]{placeins}
\usepackage{textcomp}
\usepackage{amsmath,amsfonts,amssymb}
\usepackage{graphicx}
\usepackage[colorlinks=true, allcolors=blue]{hyperref}
\usepackage{subfig}
\usepackage{graphicx}
\usepackage{mathtools, siunitx}
\usepackage{indentfirst}
\usepackage{textcomp}
\usepackage{caption}

\title{Deformable mirror-based pupil chopping for exoplanet imaging and adaptive optics}
\author[1]{Javier Perez Soto}
\author[2]{Cesar Laguna}
\author[2]{Benjamin L. Gerard}
\author[1]{Anne Dattilo}
\author[1]{Vincent Chambouleyron}
\author[1]{Rebecca Jensen-Clem}
\affil[1]{University of California, Santa Cruz}
\affil[2]{Lawrence Livermore National Laboratory}

\date{August 2023}
\authorinfo{Further author information: Send correspondence to Javer Perez Soto, jperezso@ucsc.edu}

\begin{document}

\maketitle

\section{Abstract}
Due to turbulence in the atmosphere images taken from ground-based telescopes become distorted. With adaptive optics (AO) images can be given greater clarity allowing for better observations with existing telescopes and are essential for ground-based coronagraphic exoplanet imaging instruments. A disadvantage to many AO systems is that they use sensors that can not correct for non-common path aberrations. We have developed a new focal plane wavefront sensing technique to address this problem called deformable mirror (DM)-based pupil chopping. The process involves a coronagraphic or non-coronagraphic science image and a deformable mirror, which modulates the phase by applying a local tip/tilt every other frame which enables correcting for leftover aberrations in the wavefront after a conventional AO correction. We validate this technique with both simulations (for coronagraphic and non-coronagraphic images) and testing (for non-coronagraphic images) on UCSC's Santa Cruz Extreme AO Laboratory (SEAL) testbed. We demonstrate that with as low as 250 nm of DM stroke to apply the local tip/tilt this wavefront sensor is linear for low-order Zernike modes and enables real-time control, in principle up to kHz speeds to correct for residual atmospheric turbulence.

\section{Introduction}
Adaptive optics have become an indispensable tool in ground based astronomy. Through the use of AO, diffraction-limited astronomical observations have become viable at ground level. Integral to many of these systems are the deformable mirror (DM) and wavefront sensor (WFS). Together they allow astronomers to correct wavefronts and gain better fidelity in their observations. While AO is extremely useful, other issues arise from a conventional DM+WFS AO setup. Plaguing these systems are errors called non-common path aberrations (NCPAs). NCPAs can occur for a variety of reasons such as optical components or misalignments (Ref. \citenum{hardy}). Uncorrected NCPAs are known to be a particularly major limitation to exoplanet imaging detection sensitivity (Ref. \citenum{cdi_rev}), motivating the work for this paper.

In this paper we build upon a NCPA compensation method called DM-based pupil chopping (Ref. \citenum{gerard22b}). We describe the technique in \S\ref{sec: methods}, present simulations and validating laboratory results for a non-coronagraphic setup in \S\ref{sec: non_coron}, present simulations of a coronagraphic setup in \S\ref{sec: coron}, and conclude in \S\ref{sec: conclusion}.
\section{Concept Description}
\label{sec: methods}

\begin{figure}[ht]
    \centering
    \includegraphics[scale=.55]{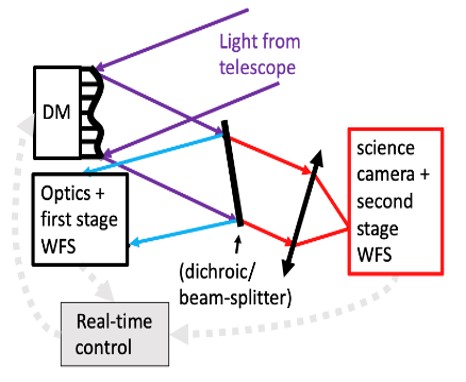}
    \caption{AO system using two wavefront sensors, from Ref. \citenum{gerard22b}. One WFS is located before the splitter and the other is at the science camera}
    \label{fig:AO system}
\end{figure}

This method incorporates two wavefront sensors rather than one. One located before a dichroic/beam splitter and the second located at the science camera (Figure \ref{fig:AO system}). 
%
The need of the second WFS comes from an action we will refer to as ``chop.'' This chop allows us to temporarily introduce a local tip/tilt aberration onto a component of the DM. An initial version of this technique was presented in Ref. \citenum{gerard22b} and \citenum{gerard23}, an optical chopper partially blocks the pupil amplitude every other frame in a pupil downstream of the splitter in Fig. \ref{fig:AO system}. However, subsequently a simpler technique was realized, presented initially in Ref. \citenum{gerard22b}, where instead a local tilt on the DM modulates the phase but effectively also modulates the pupil amplitude of the resultant downstream on-axis point spread function (PSF), which also enables a science duty cycle greater than 50\%. With this local tip/tilt applied on the DM every other frame, the secondary WFS can further correct for any remaining aberrations left by the system. 

\begin{figure}[ht]
    \centering
    \includegraphics[scale=.7]{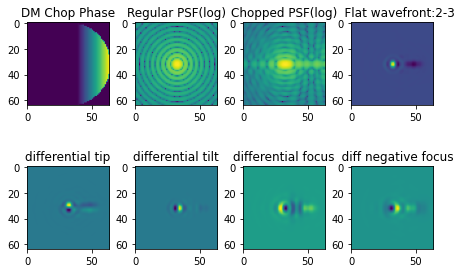}
    \caption{Adapted from Ref. \citenum{gerard22b}, DM-based pupil chopping illustration for non-coronagraphic system, using a chop fraction of 0.4 and a chop amplitude of 4 radians, and Zernike probes with amplitudes of 0.004 radians.}
    \label{fig: Non-coronagraph system}
\end{figure}

It is important to mention that the complete chopping action requires two steps. First, the DM applies a local tip/tilt to the wavefront and an obscured/chopped PSF is recorded. That local tip/tilt has two free parameters: the chop fraction---how much of the pupil to local tilt---and the chop amplitude---how much peak-to-valley amplitude to apply on the DM to make the local tip/tilt component. Second, we relax the DM and another diffraction-limited PSF (useable for science) is recorded. Then taking the difference between the chopped PSF and the regular PSF (unchopped) gives a calibrated reference image that will be referred to as ``chopping'' our image. Now that we have acquired our reference ``chop,'' subsequent wavefront reconstruction is enabled by a double difference between (1) a chopped difference pair with unknown wavefront aberrations that are the same in both the chopped and unchopped frames and (2) the above-mentioned reference chop. This double differencing process is illustrated in Fig. \ref{fig: Non-coronagraph system}, showing that the technique can resolve the focal plane sign ambiguity between positive and negative focus. 

We use a linear least-squares matrix vector multipliction wavefront reconstruction method to generate DM commands and modal coefficients from the above-described double difference focal plane images. In this paper we demonstrate this technique for low-order Zernike modes.
\section{Non-coronagraphic Simulations and Laboratory Results}
\label{sec: non_coron}
\begin{figure}[ht]
    \centering
    \includegraphics[width=0.63\textwidth]{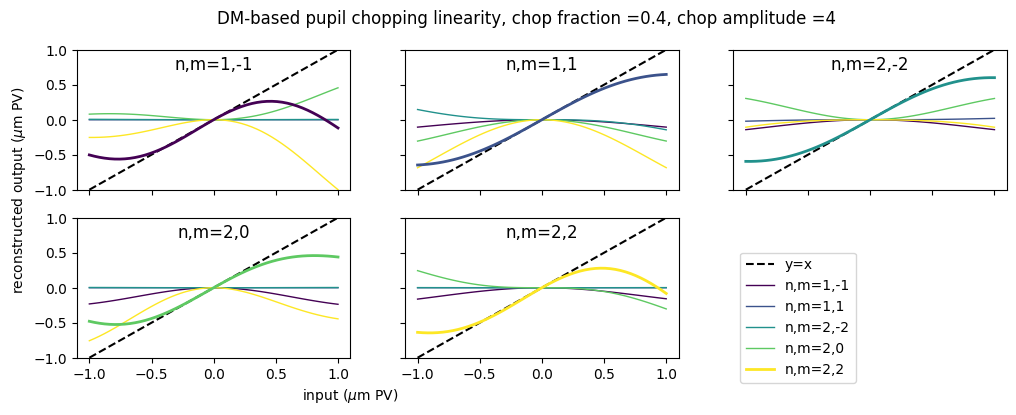}
    \includegraphics[width=0.36\textwidth]{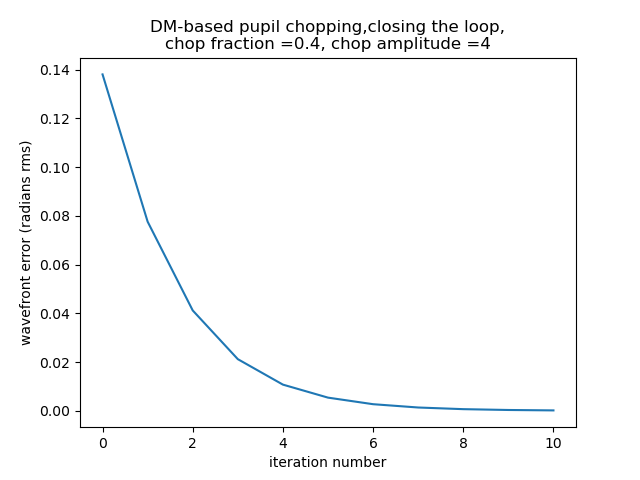}
    \caption{Simulated test bench, on the left using the chopping technique to probe linearity of different Zernike modes and on the right closing the loop (using an integrator with a gain of 1) on non-linearities for the same 5 Zernike modes.}
    \label{fig:Linearity Plots}
\end{figure}

%
In this section, building off of the work in Ref. \citenum{gerard22b}, we will describe both the simulated and laboratory test bench used to determine the feasibility of this new DM-based pupil chopping technique. 
%
%
In Fig. \ref{fig:Linearity Plots} we show simulated linearity plots for this technique, demonstrating good linear behavior in the diffraction-liminted input wavefront (i.e., $<$ 1 rad rms) regime.
For each plot, such as n,m = 1,-1, good closed-loop performance will be achieved with this technique when the given probed Zernike mode hovers around the y = x line while keeping the remaining Zernike modes at 0. Initial work in Ref. \citenum{gerard22b} used a chop amplitude of 10 rad, but good performance was found here down to a lower chop amplitude of 4 rad, which is desirable as it requires less DM stroke (corresponding to 500 nm in H band). In Fig. \ref{fig:Linearity Plots} we also show the results of closing the loop on non-linearities (i.e., simulated noiseless images with a 1 rad rms, -2 power law phase screen deprojected onto only the 5 Zernike modes measured here) for the same chop fraction and chop amplitude parameters as the linearity plots. This convergence effectively to zero wavefront error is consistent with the good linearity seen for small phase errors.



Lab data was acquired using the Santa Cruz Extreme AO Lab (SEAL; Ref. \citenum{jensen-clem21}) testbed. For the purposes of this paper, as in Ref. \citenum{gerard22b} we used a 97 actuator ALPAO DM and FLIR blackfly camera for our pupil chopping demonstration. To sufficiently flatten the wavefront to a diffraction-limited starting point for DM pupil chopping demonstrations, we initially used Shack Hartman wavefront sensor but found better performance with a non-linear pyramid wavefront sensor wavefront reconstruction algorithm (Chambouleryron et al., A\&A, in press). 

\begin{figure}[ht]
    \centering
    \includegraphics[width=0.8\textwidth]{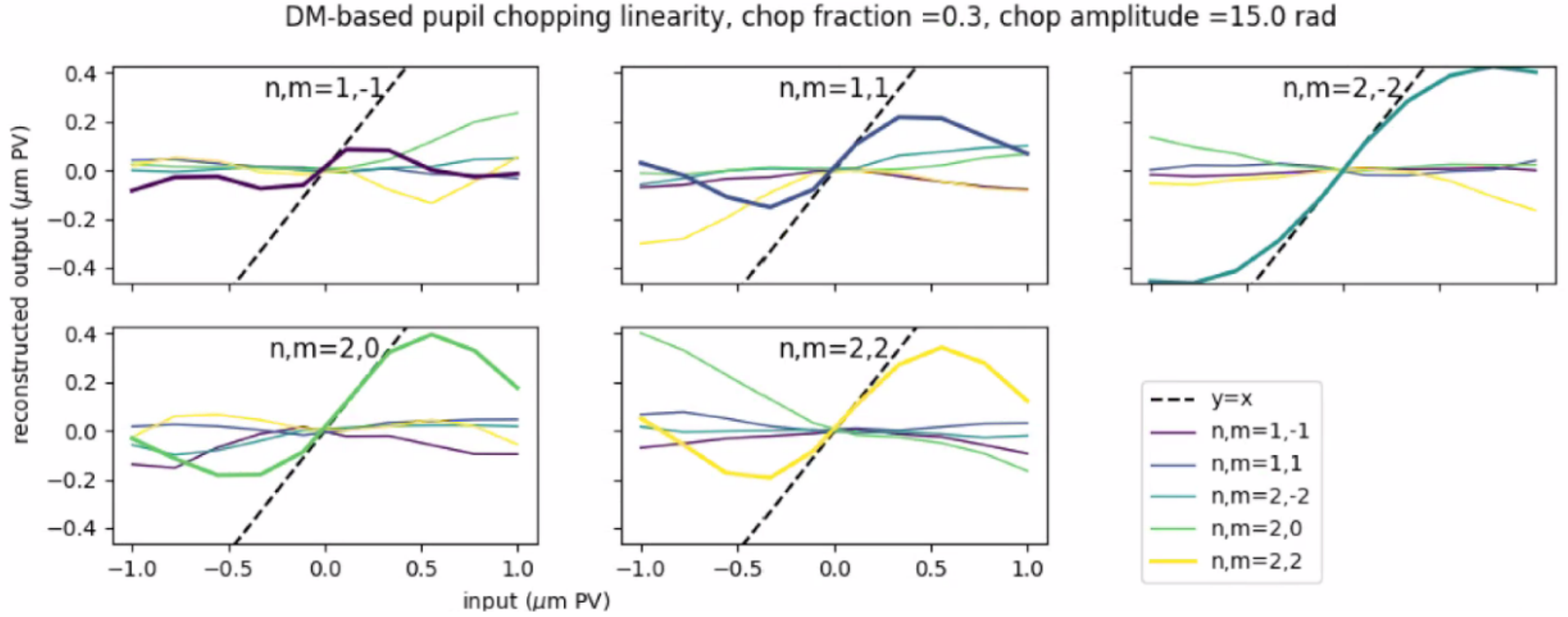}
    \caption{DM-based pupil chopping linearity plots obtained on the SEAL testbed while probing the DM with Zernike modes}
    \label{fig: PWFS Linearity Plots}
\end{figure}

With the above-presented simulations showing that the method is feasible, we then tested linearity on the SEAL test bench, shown in Fig. \ref{fig: PWFS Linearity Plots}.

These results use a chop fraction at 0.4 and chop amplitude at 15 radians. Further testing could be conducted in lowering the chop amplitude as evident from our simulations. Although simulations use amplitudes of 0.004 rad (assuming 1 nm at $\lambda$=1.6 $\mu$m), in laboratory data we used an amplitude of 1 rad out of necessity due to the limited dynamic range of the detectors, which also decreases linearity performance.
\section{Coronagraphic Simulations}
\label{sec: coron}

Coronagraphs are an attachment on telescopes that are used to block out the direct light from a bright object, such as a star, so that objects that would otherwise be hidden are now viewable (Ref. \citenum{coron}). We have already demonstrated good linearity and closed-loop convergence for a non-coronagraphic  DM pupil chopping system (\S\ref{sec: non_coron}). From here we wanted to know if applying a coronagraph to the system would improve the DM pupil chopping, as in being able to close the loops on the system with less DM stroke.

\begin{figure}[h!]
  \centering
  \subfloat{{\includegraphics[width=\textwidth]{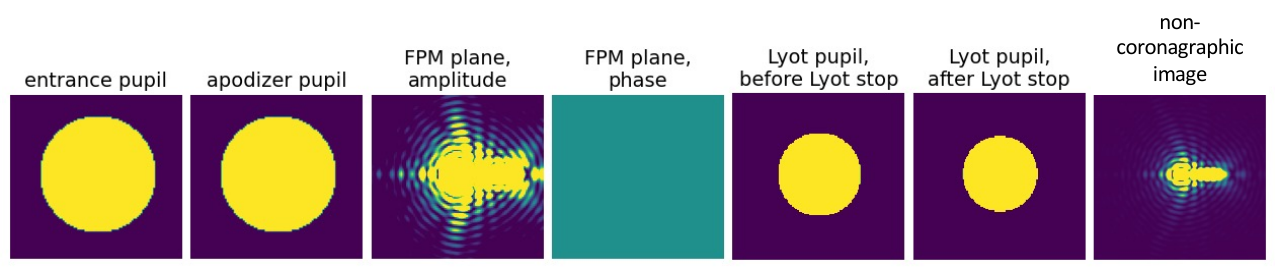}}}%
  \qquad
  \subfloat{{\includegraphics[width=\textwidth]{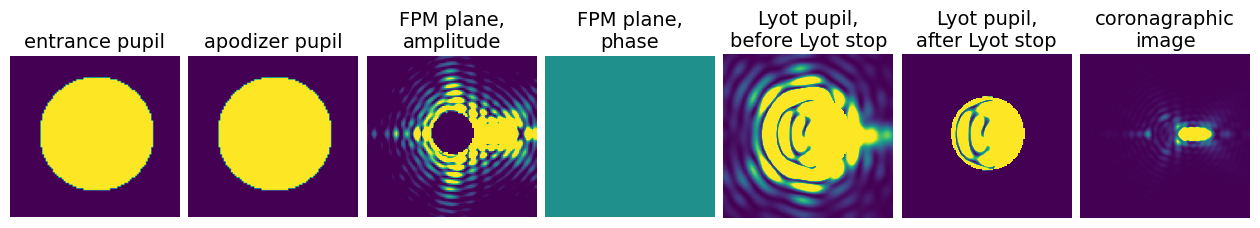}}}%
  \caption{The concept of non-coronagraphic (above) vs. coronagraphic (below) pupil chopping, using a chop fraction of 0.4 and chop amplitude of 10 rad in both cases.}
\label{fig: concept_illustration}
\end{figure}
%
\begin{figure}[ht]
  \centering
  \includegraphics[width=10cm]{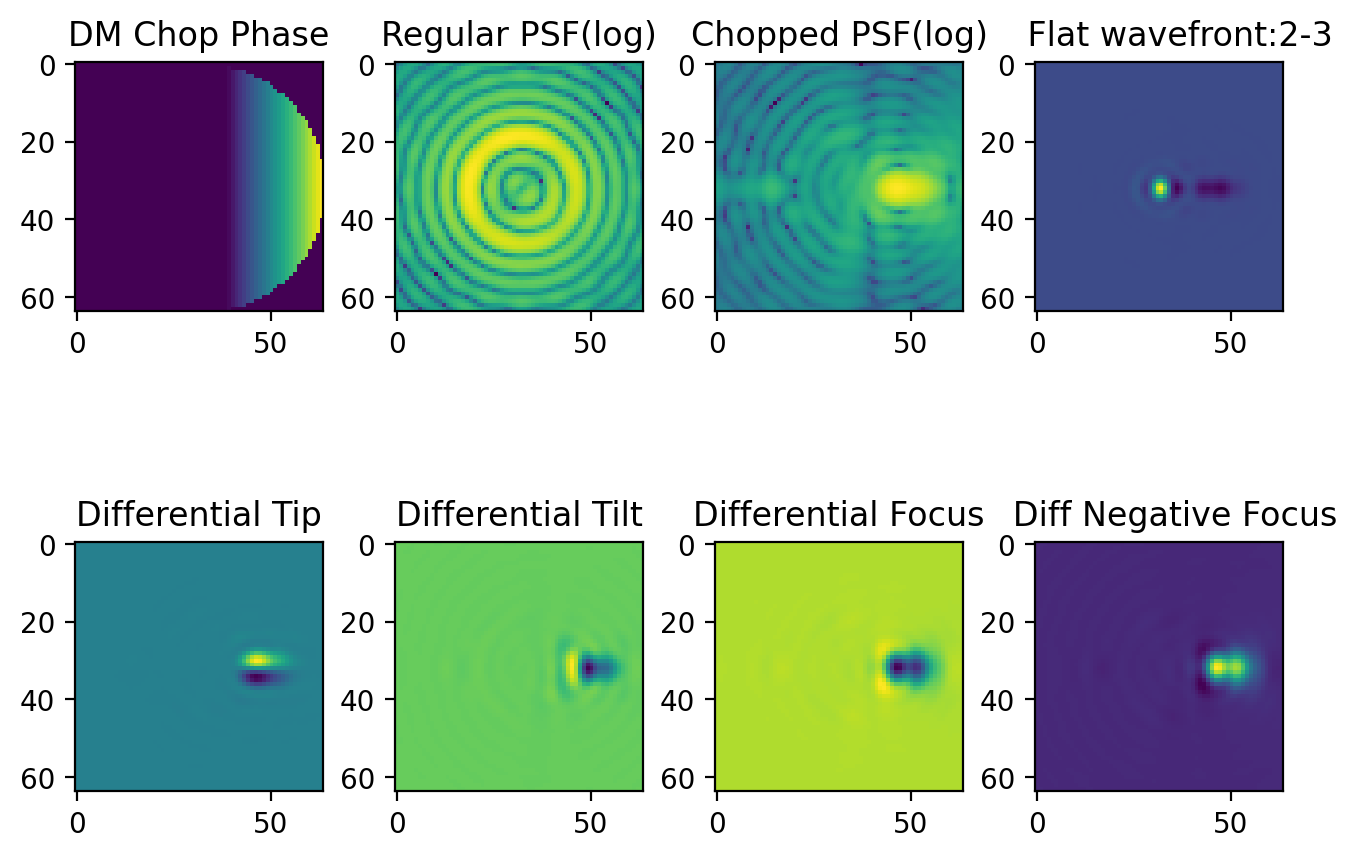}%
  \caption{An illustration of coronagraphic DM-based pupil chopping, depicting the effects a Lyot cornograph produces on the system for various Zernike modes. The Top right image is a single difference, while the bottom row shows double difference images. All images use a chop fraction of 0.4, chop amplitude of 10 rad, and Zernike probe amplitude of 0.004 rad.}
  \label{fig: System Process}
\end{figure}
To simulate a simple Lyot coronagraph we introduce two new optical elements, a focal plane mask and a Lyot stop. The coronagraph mask will block out the majority of the starlight and the Lyot stop, slightly smaller than the aperture size, will remove the effects of diffraction so that the final image is relatively clear. Fig. \ref{fig: concept_illustration} illustrates the effects of pupil chopping with vs. without a coronagraph.

We tested the feasibility of loop closure for the coronagraphic system in the same manner that we did the non-coronagraphic system, by applying low order Zernike modes. Figure \ref{fig: System Process} shows simulated low order wavefront/Zernike modes using this pupil chopping method with a coronagraph applied and illustrates the ability of this technique to resolve the sign of focus using the coronagraphic image plane.

\begin{figure}[ht]
  \centering
  \subfloat{{\includegraphics[width=11cm]{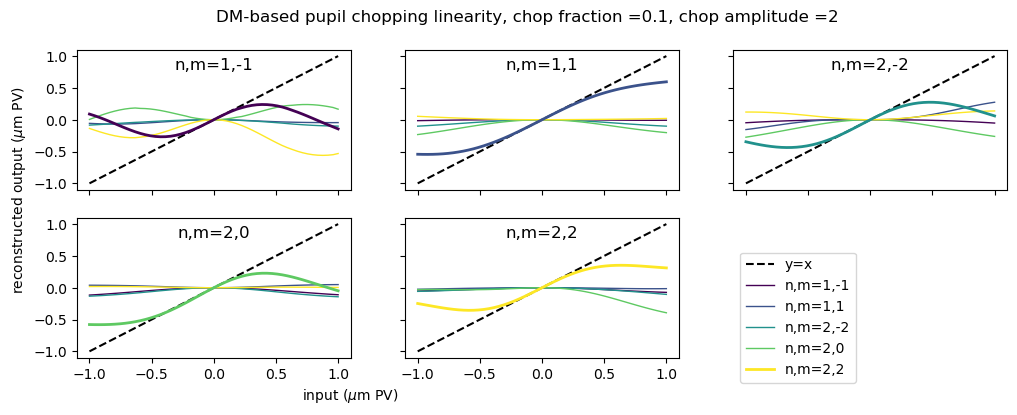}}}%
  \qquad
  \subfloat{{\includegraphics[width=5cm]{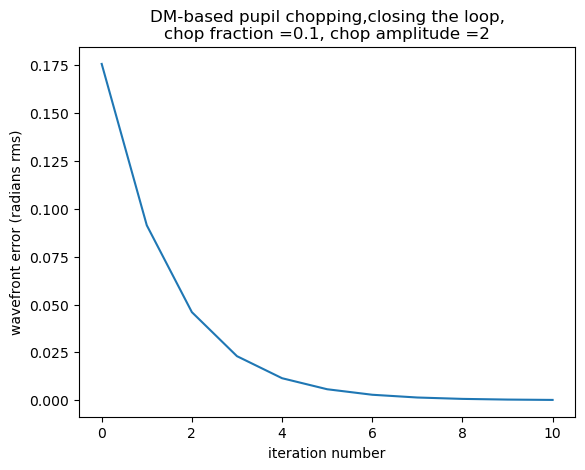}}}%
  \caption{Simulated Linearity curves and loop-closing on static non-linearities (analagous to Fig. \ref{fig:Linearity Plots} for a non-coronagraphic system). If linearity is possible in the system we will be able to close the loops. Less iterations to converge in closing the loops corresponds to lower non-linearities in the error budget.}
\label{fig:linearity good}
\end{figure}

Fig. \ref{fig:linearity good} shows the low DM stroke needed to enable good linearity.
In H band a 2 rad local tilt as in Fig. \ref{fig:linearity good} would use a stroke of 250 nm, showing significant improvement of the analagous 500 $\mu$m stroke requirement for non-coronagraphic pupil chopping.

\begin{figure}[ht]
  \centering
  \subfloat{{\includegraphics[width=11cm]{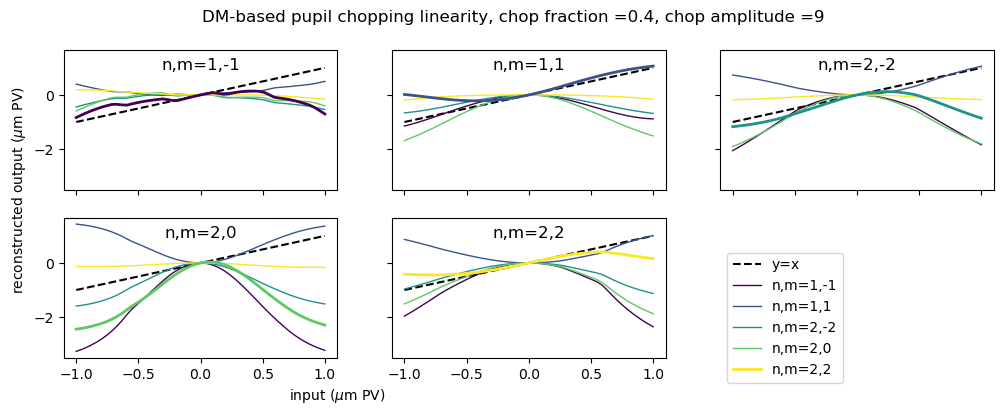}}}%
  \qquad
  \subfloat{{\includegraphics[width=5cm]{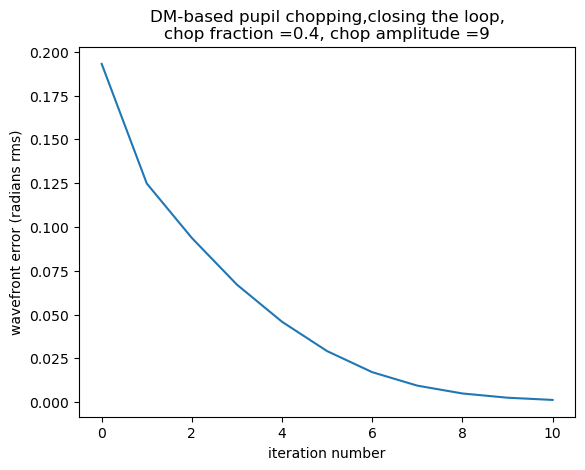}}}%
  \caption{An example of chop fraction and amplitude parameters that produce less optimal linearity as compared to Fig. \ref{fig:linearity good}. In this case, the system does close the loops but it takes more iterations than in Fig. \ref{fig:linearity bad}, indicating greater non-linearities in the wavefront error budget with this WFS.}
\label{fig:linearity bad}
\end{figure}

\begin{figure}[ht]
  \centering
  \subfloat{{\includegraphics[width=6cm]{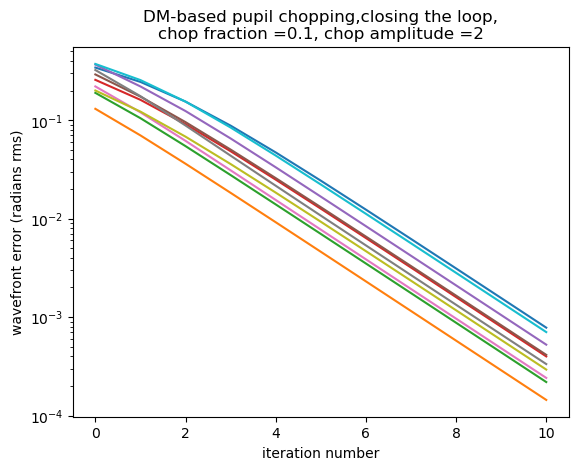}}}%
  \qquad
  \subfloat{{\includegraphics[width=6cm]{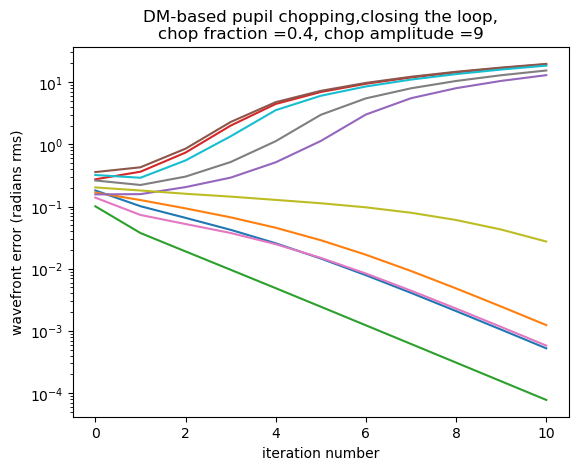}}}%
  \caption{Figure on left shows total convergence whereas the figure on the right has some iterations that converge while others diverge. Those that diverged at least half the time were deemed to have insufficient chop fractions and/or amplitudes.}
\label{fig:iterations}
\end{figure}

Interestingly, when testing various chop fractions and amplitudes certain values would provide loop-closing convergence plots that would suggest that loop closure is possible, yet their linearity plots disagree such as in figure \ref{fig:linearity bad}.
\pagebreak{}
When deciding the cut-off for chop fractions and amplitudes we did 10 iterations (in each iteration closing the loop on a random realization of wavefront error but with each normalized to 1 rad rms projected onto the 5 Zernike modes considered in this paper) of each value in order to determine whether or not the loops would close as shown in Figure \ref{fig:iterations}. If the loops closed for all 10 iterations we know that the possibility of loop closure is good. However, if the loops closed for only a portion of the 10 iterations we decided it would be best to choose other chop fractions and amplitudes.

\section{Conclusion}
\label{sec: conclusion}
In this manuscript we have built on the work in Ref. \citenum{gerard22b} in the following ways:
\begin{enumerate}
    \item\label{pt: one} We have demonstrated that non-coronagraphic DM-based pupil chopping---a concept that involves acquiring two non-coronagraphic PSF images where the DM in one of the two applies a local chopping pattern---enables good linear reconstruction for low order Zernike modes down to about 500 nm of DM stroke in H band, which is more than a 2$\times$ reduction compared to the findings in Ref. \citenum{gerard22b}.
    \item We have also validated simulations of point \ref{pt: one} by presenting laboratory results with the SEAL testbed, also showing good linearity.
    \item We have simulated DM-based pupil chopping with a coronagraphic system, showing that DM stroke can be further reduced by $\sim2\times$ compared to a non-coronagraphic system without compromising the performance of this technique.
\end{enumerate}
Future work on this topic will aim to simulate the linearity of this technique for higher order modes, demonstrate closed-loop laboratory performance for a non-coroangraphic system, simulate this technique with additional small inner-working angle coronagraphs, and demonstrate closed-loop performance for a laboratory coronagraphic system.
\section*{Acknowledgements}

We would like to extend our gratitude to the University of California, Santa Cruz, the entire SEAL lab team and UCSC's LAMAT NSF REU. Author J. Perez Soto thanks Dr. Benjamin L. Gerard and Deana Tanguay for their support and encouragement throughout the entire process. This work was performed under the auspices of the U.S. Department of Energy by Lawrence Livermore National Laboratory under Contract DE-AC52-07NA27344. This document number is LLNL-PROC-853110.

\bibliography{report} 

\begin{thebibliography}{1}

\bibitem{hardy}
{Hardy}, J.~W.,  [{\em {Adaptive Optics for Astronomical
  Telescopes}}{\nolinebreak\hspace{0.1em}]} (1998).

\bibitem{cdi_rev}
{Jovanovic}, N., {Absil}, O., {Baudoz}, P., {Beaulieu}, M., {Bottom}, M.,
  {Cady}, E., {Carlomagno}, B., {Carlotti}, A., {Doelman}, D., {Fogarty}, K.,
  {Galicher}, R., {Guyon}, O., {Haffert}, S., {Huby}, E., {Jewell}, J.,
  {Keller}, C., {Kenworthy}, M.~A., {Knight}, J., {K{\"u}hn}, J., {Miller}, K.,
  {Mazoyer}, J., {N'Diaye}, M., {Por}, E., {Pueyo}, L., {Riggs}, A.~J.~E.,
  {Ruane}, G., {Sirbu}, D., {Snik}, F., {Wallace}, J.~K., {Wilby}, M., and
  {Ygouf}, M., ``{Review of high-contrast imaging systems for current and
  future ground-based and space-based telescopes: Part II. Common path
  wavefront sensing/control and coherent differential imaging},'' in [{\em
  Adaptive Optics Systems VI}{\nolinebreak\hspace{0.1em}]},  {Close}, L.~M.,
  {Schreiber}, L., and {Schmidt}, D., eds., {\em Society of Photo-Optical
  Instrumentation Engineers (SPIE) Conference Series} {\bf 10703},  107031U
  (July 2018).

\bibitem{gerard22b}
{Gerard}, B.~L., {Perez-Soto}, J., {Chambouleyron}, V., {van Kooten}, M. A.~M.,
  {Dillon}, D., {Cetre}, S., {Jensen-Clem}, R., {Fu}, Q., {Amata}, H., and
  {Heidrich}, W., ``{Various wavefront sensing and control developments on the
  Santa Cruz Extreme AO Laboratory (SEAL) testbed},'' in [{\em Adaptive Optics
  Systems VIII}{\nolinebreak\hspace{0.1em}]},  {Schreiber}, L., {Schmidt}, D.,
  and {Vernet}, E., eds., {\em Society of Photo-Optical Instrumentation
  Engineers (SPIE) Conference Series} {\bf 12185},  121852H (Aug. 2022).

\bibitem{gerard23}
{Gerard}, B.~L., {Dillon}, D., {Cetre}, S., and {Jensen-Clem}, R.,
  ``{High-speed Focal Plane Wave Front Sensing with an Optical Chopper},'' {\em
  Publications of the Astronomical Society of the Pacific}~{\bf 135},  024502
  (Feb. 2023).

\bibitem{jensen-clem21}
{Jensen-Clem}, R., {Dillon}, D., {Gerard}, B., {van Kooten}, M.~A.~M.,
  {Fowler}, J., {Kupke}, R., {Cetre}, S., {Sanchez}, D., {Hinz}, P., {Laguna},
  C., {Doelman}, D., and {Snik}, F., ``{The Santa Cruz Extreme AO Lab (SEAL):
  design and first light},'' in [{\em Society of Photo-Optical Instrumentation
  Engineers (SPIE) Conference Series}{\nolinebreak\hspace{0.1em}]},  {\em
  Society of Photo-Optical Instrumentation Engineers (SPIE) Conference Series}
  {\bf 11823},  118231D (Sept. 2021).

\bibitem{coron}
{Ruane}, G., {Riggs}, A., {Mazoyer}, J., {Por}, E.~H., {N'Diaye}, M., {Huby},
  E., {Baudoz}, P., {Galicher}, R., {Douglas}, E., {Knight}, J., {Carlomagno},
  B., {Fogarty}, K., {Pueyo}, L., {Zimmerman}, N., {Absil}, O., {Beaulieu}, M.,
  {Cady}, E., {Carlotti}, A., {Doelman}, D., {Guyon}, O., {Haffert}, S.,
  {Jewell}, J., {Jovanovic}, N., {Keller}, C., {Kenworthy}, M.~A., {Kuhn}, J.,
  {Miller}, K., {Sirbu}, D., {Snik}, F., {Wallace}, J.~K., {Wilby}, M., and
  {Ygouf}, M., ``{Review of high-contrast imaging systems for current and
  future ground- and space-based telescopes I: coronagraph design methods and
  optical performance metrics},'' in [{\em Space Telescopes and Instrumentation
  2018: Optical, Infrared, and Millimeter Wave}{\nolinebreak\hspace{0.1em}]},
  {Lystrup}, M., {MacEwen}, H.~A., {Fazio}, G.~G., {Batalha}, N., {Siegler},
  N., and {Tong}, E.~C., eds., {\em Society of Photo-Optical Instrumentation
  Engineers (SPIE) Conference Series} {\bf 10698},  106982S (Aug. 2018).

\end{thebibliography}
\bibliographystyle{spiebib} 

\end{document}